\documentclass[conference]{sig-alternate-10pt}

\usepackage{xcolor}

\usepackage{colortbl}
\usepackage{multirow,bigdelim}
\usepackage{arydshln}

\usepackage{graphicx}
\usepackage{dcolumn}
\usepackage{bm}
\usepackage{epsfig}
\usepackage{psfrag}
\usepackage{subfigure}
\usepackage{color}
\usepackage{amsmath, amssymb, latexsym,amsfonts,epsfig, graphicx, verbatim}
\usepackage[nospace]{cite}
\usepackage{verbatim}
\usepackage{xspace}

\newcommand{\eqn}[1]{Eq.(\ref{#1})}
\newcommand{\Sec}[1]{Sec.~\ref{#1}}
\newcommand{\Fig}[1]{Fig.~\ref{#1}}
\newcommand{\Tab}[1]{Table~\ref{#1}}

\newcommand{\Mean}[1]{\mathbb{E}\left[#1\right]}

\newcommand{\Angles}[1]{\langle #1\rangle}

\newcommand{\eq}{\!\!=\!}
\newcommand{\m}{\!-\!}

\newcommand{\w}{\textrm{w}}

\newcommand{\unique}{{\scriptscriptstyle\textrm{uniq}}}
\newcommand{\UIS}{{\scriptscriptstyle\textrm{UIS}}}
\newcommand{\WIS}{{\scriptscriptstyle\textrm{WIS}}}
\newcommand{\RW}{{\scriptscriptstyle\textrm{RW}}}

\newcommand{\est}[1]{\widehat{#1}}

\newcommand{\ie}{{\em i.e., }}
\newcommand{\eg}{{\em e.g., }}

\newcommand{\NODE}{\textrm{NODE}\xspace}

\newcommand{\SimpleThinning}{SimpleThinning\xspace}
\newcommand{\ShiftedThinning}{ShiftedThinning\xspace}
\newcommand{\Margin}{SafetyMargin\xspace}
\newcommand{\col}{{\scriptscriptstyle\textrm{col}}}
\newcommand{\xcol}{{\scriptscriptstyle\textrm{xcol}}}
\newcommand{\IND}{\textrm{IE}\xspace}
\newcommand{\INDA}{\textrm{IE1}\xspace}
\newcommand{\INDB}{\textrm{IE2}\xspace}
\newcommand{\ind}{{\scriptscriptstyle\textrm{IE}}}
\newcommand{\indA}{{\scriptscriptstyle\textrm{IE1}}}
\newcommand{\indB}{{\scriptscriptstyle\textrm{IE2}}}

\definecolor{Gray}{gray}{0.9}

\begin{document}

\title{Graph Size Estimation}

\numberofauthors{3}


\author{
%
%
\alignauthor
Maciej Kurant\\
       \affaddr{ETHZ}\\
       \email{maciej.kurant@tik.ee.ethz.ch}
\alignauthor Carter T. Butts\\
       \affaddr{UC Irvine}\\
       \email{buttsc@uci.edu}
\and  
\alignauthor Athina Markopoulou\\
       \affaddr{UC Irvine}\\
       \email{athina@uci.edu}
}

\maketitle

\begin{abstract}

Many online networks are not fully known and are often studied via sampling.  Random  Walk (RW) based techniques are the current state-of-the-art for estimating nodal attributes and local graph properties, but estimating global properties remains a challenge.  In this paper, we are interested in a fundamental property of this type --- the graph size~$N$, \ie the number of its nodes. Existing methods for estimating~$N$ are (i)~inefficient and (ii)~cannot be easily used with RW sampling due to dependence between successive samples. In this paper, we address both problems. 

First, we propose \IND (Induced Edges), an efficient technique for estimating $N$ from an independence sample of graph's nodes. \IND exploits the edges induced on the sampled nodes. 
Second, we introduce \Margin, a method that corrects estimators for dependence in RW samples.  
Finally, we combine these two stand-alone techniques to obtain a RW-based graph size estimator. 
We evaluate our approach in simulations on a wide range of real-life topologies, and on several samples of Facebook. 
\IND with \Margin typically requires at least 10 times fewer samples than the state-of-the-art techniques (over 100 times in the case of Facebook) for the same estimation error.

\end{abstract}

\section{Introduction}

An important and fundamental graph property is its \emph{size}, \ie number of nodes $N$. 
This is a property with practical as well as theoretical importance.  For example, consider the market value (\eg, stock price) of an online social network (OSN) service provider such as Facebook. 
Among the criteria analysts would consider when valuing such a firm is its current number of users~$N$, and 
its growth rate (\ie new users per month). 
These numbers are critical not only for investors, but also for various business decisions, such as choosing the medium for an advertising campaign, or for lunching a social application. 

In some cases, OSN providers officially publish the total number of users. However, these numbers may be
(i)~outdated, 
(ii)~incorrect (there exist strong incentives to report large $N$), or
(iii)~difficult to compare across networks (\eg Facebook publishes the number of its \emph{active} users, which is different from the total published by its competitors).

In other cases, the graph size may be not available at all. 
For example, in \emph{distributed computer systems}, the entities (\eg nodes in a P2P network) have only a local view of the system (a list of neighbors). 
While this often leads to better scalability and reliability, it makes it much harder to obtain system parameters that are trivially known in a centralized architecture. 
One of them is the system size~$N$ - a common input parameter in various distributed protocols, such as overlay maintenance~\cite{Malkhi2002} or 
routing~\cite{Eugster2003}. 



$N$~is also typically unknown in the study of \emph{online media}. 
WWW, blogging platforms, instant messaging and OSNs are all rich in information content contributed by millions of individuals and organizations. 
The knowledge of the structure and the processes in these information networks can be used to track the spread of memes (news, topics, ideas, URLs)~\cite{Leskovec2009a,Galuba2010},
predict the outcome of presidential elections~\cite{Ackland2004,Adamic2005}, 
or improve a marketing campaign~\cite{kleinberg-book}. 
Unfortunately, complete social media data is often impossible to collect, and the results obtained on incomplete datasets are potentially biased~\cite{deChoudhury2011}. 
To assess the completeness of collected data (and thus the extent of this bias), one can compare the size of the sampled part with the estimated total size~$N$ of the information network.


Finally, estimating the size of hidden populations such as drug users or HIV positives is a major challenge in the \emph{social sciences}. 
One of the main sampling techniques currently used in this context is a variation of RW called Respondent-Driven Sampling (RDS)~\cite{Heckathorn97_RDS_introduction,Salganik2004,Rasti09-RDS,Goel2010}. 
``RDS is now widely used in the public health community and has been recently applied in more than 120 studies in more than 20 countries, involving a total of more than 32000 participants''~\cite{Goel2010}. 
Because these field studies have a significant monetary cost, any improvement in measurement efficiency directly leads to concrete budget savings.

In all these cases, it is highly desired to have an efficient way of estimating the size of a graph based on sampled data. 
One of the most popular---and often the only feasible in practice (see Footnote~\ref{footnote:UIS})---sampling technique is Random Walk~(RW). 
RW-based sampling has been used to sample the WWW~\cite{Henzinger2000}, P2P networks~\cite{Stutzbach2006-unbiased-p2p,Rasti09-RDS, Gkantsidis2004}, 
OSNs \cite{Twitter08,Rasti2008,Gjoka2010,Mohaisen2010}, 
and ``offline'' social networks~\cite{Heckathorn97_RDS_introduction,Salganik2004,Rasti09-RDS,Goel2010}. 
In this paper, we focus on and derive \emph{efficient and practical RW-based estimators of graph size}~$N$. 
Our contributions are the following. 

\begin{figure}
\psfrag{A}[l][c][1]{a) \NODE (state of the art)}
\psfrag{B}[l][c][1]{b) \IND (this paper)}
\psfrag{L4}[l][c][0.8]{Induced edges (\IND)}
\psfrag{L3}[l][c][0.8]{Sampled nodes}
\psfrag{L2}[l][c][0.8]{Sampled nodes}
\psfrag{L0}[l][c][0.8]{Unsampled nodes}
\centering
\includegraphics[width=0.5\textwidth]{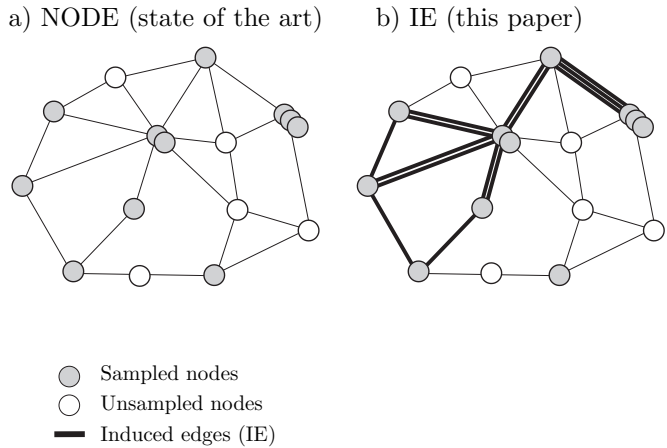}
\caption{
Two families of techniques to estimate the graph size~$N$ based on a (uniform or non-uniform) sample of nodes, with replacements. 
\quad (a)~\NODE techniques (state of the art) exploit node collisions (repetitions). 
Here, among $n\eq 11$ sampled nodes, we observe $n^\unique\eq 8$ unique nodes, and $n^\col\eq4$ node collisions; \NODE uses these numbers to estimate~$N$. 
\quad (b)~\IND techniques (introduced in this paper) exploit the edges induced on the sampled nodes (with repetitions). Here, we have $n^\ind=14$ such edges. 
}
\label{fig:NODEvsIND}
\end{figure}

First, in~\Sec{sec:Graph size estimation based on induced subgraph}, we propose \IND (for \emph{Induced Edges}), a family of efficient techniques to estimate~$N$ based on an independence sample (uniform or not) of its nodes. 
\IND exploits the number of edges induced on the sampled nodes (see~\Fig{fig:NODEvsIND}(b)), which is fundamentally different from the state-of-the-art techniques that exploit the node repetitions within the sample (see~\Fig{fig:NODEvsIND}(a) and~\Sec{sec:Related work}).


Second, in \Sec{sec:random_walk}, we extend \IND to accept RW samples. Here, the main challenge lies in that the consecutive RW samples are strongly \emph{dependent}, which critically impacts the estimation results. 
We address this problem by introducing several RW dependence reduction techniques, including \Margin \ - our best performer. \Margin is a stand-alone technique that can be applied in other RW-based estimation problems as well. 

Third, in~\Sec{sec:Implementation issues}, we discuss the practical implementation issues related to our estimators, and make our  
efficient \texttt{python} implementation available at \cite{WWW_N_est}. 

Fourth, in \Sec{sec:results}, we evaluate our approach in simulations on a wide range of real-life topologies and on several samples of Facebook (confirming the officially announced numbers). Compared to the state-of-the-art solutions, we typically observe several-fold gain in sampling cost for each \IND and \Margin, separately. When combined together, \IND with \Margin usually requires at least 10 times fewer samples versus standard methods (over 100 times in the case of Facebook) for the same estimation error.

\section{Notation}\label{sec:Notation}

Let~$G=(V,E)$ be an undirected, connected graph. 
Let \emph{graph size} $N\eq|V|$ be the number of nodes in~$G$. 
Our goal in this paper is to estimate~$N$,\footnote{If needed, given the estimated~$N$, one can easily estimate the  number of edges as~$|E|=N\cdot\Angles{k}$/2, where the average node degree $\Angles{k}$ can be calculated as in \eqn{eq:av_degree_est_uis} or \eqn{eq:av_degree_est_wis}.} based on a sample~$S=[s_1,s_2,\ldots s_n]$ of $n\eq|S|$ nodes, with replacements. For every sampled node~$s\in S$, we know the list of its neighbors $\mathcal{N}(s)$. 
Depending on the way $S$ is collected, we distinguish the following sampling techniques:
\begin{itemize}
	\item \emph{Uniform Independence Sample} (UIS): The nodes in~$S$ are collected independently, uniformly at random, with replacement.\footnote{\label{footnote:UIS}Collecting a UIS node sample would be a trivial task if we had a list of all nodes in the graph. But then, of course, the graphs size needs no estimation - we know it precisely. 
Alternatively, a UIS sample can be sometimes obtained by rejection sampling of the userID space~\cite{Gjoka2010}. 
However, too large userID space, such as 64-bit space currently used by Facebook, makes this approach completely impractical (unless some additional features can be exploited, as in \cite{Rejaie2010,Zhou2011}). 
}
	\item \emph{Weighted Independence Sample} (WIS): Every node $v\in S$ has a sampling probability proportional to its weight~$\w(v)$. The nodes are sampled with replacement.

	\item \emph{Random Walk} (RW): At every iteration, the next node is chosen uniformly at random from all neighbors of the current node.\footnote{Although all our results also apply directly to Weighted Random Walks~\cite{Kurant2011_SWRW}, for notation simplicity we limit the presentation to simple (unweighted) RWs.}

\end{itemize}

\section{Related Work (NODE)}\label{sec:Related work}



Population size estimation has a long history. 
Most of the existing size estimation techniques are based on \emph{node repetitions} in the collected sample~$S$. We refer to these techniques collectively as \NODE, and illustrate them in \Fig{fig:NODEvsIND}(a).

\subsection{UIS - Uniform Independence Sample}
Under UIS, 
there exist several existing approaches to estimate the population size.  The most prominent are the following.

\subsubsection{Capture-Recapture}

In this classic method~\cite{Sekar1949,Shapiro1950, Good1950}, we independently collect two uniform samples $S^\unique_1$ and $S^\unique_2$, without replacement. 
The population size can then be estimated by (or by variations of):
\begin{equation}\label{eq:capture}
	\est{N}_{\UIS}^{\scriptscriptstyle\textrm{C-R}} \ =\  \frac{|S^\unique_1|\cdot |S^\unique_2|}{|S^\unique_1\cap S^\unique_2|}.
\end{equation}
We can apply \eqn{eq:capture} to a UIS sample~$S$ by randomly splitting $S$ into two equal-sized subsamples $S_1$ and $S_2$, and then discarding the repetitions within each of them to obtain $S^\unique_2$ and $S^\unique_1$. 

Note that in this last step, we discard potentially valuable information, which may limit the performance of the estimator~\eqn{eq:capture}. Below, we present techniques better suited to a UIS sample.

\subsubsection{Unique Element Counting}
%

Population size estimation can be mapped to the problem of estimating the number of species in biology, where every node in~$V$ is a separate species (see \cite{Bunge1993} for a good review). \cite{Good1950}, page 73, uses maximum likelihood estimation (MLE) to derive the following approximation:
\begin{equation}\label{eq:est(N) UNI unique}
	n^\unique \ \cong \ N(1-e^{-n/N}),	
\end{equation}
where $n^\unique$ is the number of seen species. In our context, $n^\unique$ is the number of unique nodes in~$S$. For example, in \Fig{fig:NODEvsIND}(a), $n^\unique=8$.  
Because~$N$ is the only unknown, we can solve \eqn{eq:est(N) UNI unique} and obtain an estimate~$\est{N}$ of the population size~$N$. 

An exact version of this MLE estimator was first given in~\cite{Driml1967} defining~$\est{N}_{\UIS}$ as the smallest integer~$N\geq n^\unique$ that satisfies
\begin{equation}
	\frac{N+1}{N+1-n^\unique} \left(\frac{N}{N+1}\right)^n \ < \ 1.
\end{equation}
Its uniqueness was shown in~\cite{Finkelstein1998}. \cite{Ye2010b} proposed an efficient method of evaluating it. 

\subsubsection{Collision Counting}
Another approach is to 
study the number~$n^\col$ of \emph{collisions} in the sample~\cite{Bawa2004,Massoulie2006,Katzir2011}. 
A collision is a pair of identical samples. More precisely,
\footnote{In the sum in \eqn{eq: n^col} and in many equations that follow, indexes $i,j$ run from 1 to $n$, \ie across the entire sample~$S$.} 
\begin{equation}\label{eq: n^col}
	n^\col = \sum_{i<j} 1_{\{s_i=s_j\}}.
\end{equation}
For example, in \Fig{fig:NODEvsIND}(a), $n^\col=4$. 
Note that we usually have $n^\col + n^\unique \neq n$. 
%
We can now estimate the population size~$N$ by~\cite{Katzir2011}
\begin{equation}\label{eq:est(N) UNI collisions}
	\est{N}_{\UIS}^{\scriptscriptstyle\textrm{NODE}} \ = \ \frac{n^2}{n^\col}.
\end{equation}
%


%
%

\subsection{WIS - Weighted Independence Sample}\label{subsec:WIS - Weighted Independence Sample}

\cite{Katzir2011} provides an elegant extension of the estimator \eqn{eq:est(N) UNI collisions} to cover the WIS case, as follows:\footnote{Strictly speaking, \cite{Katzir2011} considers only the case with node degrees serving as node weights, \ie with $\w(v)=\deg(v)$. The more general version given here trivially follows from~\cite{Katzir2011}.}
\begin{eqnarray}\label{eq:est(N) WIS}
\label{eq:est_N WWW}	\est{N}_{\WIS}^{\scriptscriptstyle\textrm{NODE}}  &=& \ \frac{\displaystyle \sum_{s\in S} \w(s) \ \cdot \ \sum_{s\in S} \frac{1}{\w(s)}}{n^\col}.
\end{eqnarray}
Under UIS, \eqn{eq:est(N) WIS} reduces to \eqn{eq:est(N) UNI collisions}. 
However, it was shown in~\cite{Katzir2011} that \eqn{eq:est(N) WIS} under WIS typically outperforms \eqn{eq:est(N) UNI collisions} under UIS. 

\subsection{Other Approaches}

Various other approaches to the size estimation problem exist, but are not presented here because they: (i)~depend on special features of the network or service being studied, and are not broadly applicable; (ii)~depend on case-specific knowledge of the network or service (again, limiting applicability); and/or (iii)~are less efficient than the approaches described above.  Among the more prominent of these are the following.  

\paragraph{ Random Walk (RW) Tours} This is a family of techniques~\cite{Marchetti-Spaccamela1989, Massoulie2006,Ye2010b} that perform RWs until they return to the starting node.
However, these approaches are very inefficient, as they often require a sample comparable with the graph size~\cite{Massoulie2006,Ye2010b,Katzir2011}, and therefore we do not consider them in this paper. 

\paragraph{Traceroute}
The Internet at the IP layer gives us another widely used sampling method, \texttt{traceroute}, which can be interpreted as a rough approximation of a shortest path between two nodes. \cite{Viger2007,Kolaczyk2009} propose two graph size estimators that take as input a \texttt{traceroute} sample. 

\paragraph{Model-based Estimation}
Finally, one can assume something about the distribution of involved variables, which leads to a model-based estimation. Recently, \cite{Naini2011} used such an approach to estimate the number of Bluetooth devices in an enclosed area. 

%
\section{Induced Edge (IE) Techniques}
\label{sec:Graph size estimation based on induced subgraph}

In this paper, we take an approach that is fundamentally different from the state-of-the-art \NODE family of techniques described in~\Sec{sec:Related work} and \Fig{fig:NODEvsIND}(a). We consider not only the sampled set~$S$ of nodes, but also their sampled\footnote{In theory, one could also consider non-sampled neighbors, which is sometimes referred to as \emph{star}~\cite{Kolaczyk2009} or \emph{social}~\cite{Kumar2012} sampling. However, except for some special cases~\cite{Kurant2011_SWRW,Kurant2012}, the resulting estimators would require the knowledge of the sampling weights (degrees) of non-sampled nodes~\cite{Kumar2012}, which is rarely available in practice. }
 neighbors. In other words, we 
study the edges of $G$ \emph{induced on} $S$, as illustrated in \Fig{fig:NODEvsIND}(b). We therefore refer to this family of techniques collectively as \IND (Induced Edges). 
Under \IND, we observe an edge $\{u,v\}$ only when $u,v\in S$, \ie when both its end-nodes are sampled. 
Let $n^\ind$ be the number of such edges, \ie
\begin{eqnarray}\label{eq:n^ind}
	n^\ind &=& \sum_{i<j} 1_{\{\{s_i,s_j\} \in E\}}. 
\end{eqnarray}
Note that if nodes are repeated in $S$, we count every occurrence of each node separately. For example, in \Fig{fig:NODEvsIND}(b), $n^\ind=14$. 

Intuitively, \IND has large potential, especially in dense graphs. 
Indeed, in a particular iteration of UIS, node~$v$ is re-sampled with probability equal to $1/N$, but one of $v$'s neighbors is sampled with probability $\deg(v)/N$. Consequently, \IND observes $\Angles{k}$ (average node degree) times more collisions than \NODE, which, for typical online graphs such as Facebook with $\Angles{k}=150+$, provides far more information to exploit.

Below, we develop two approaches that exploit \IND: \INDA and \INDB. 
In their basic forms, they accept as input independence node samples (UIS or WIS); we extend them to RW samples in \Sec{sec:random_walk}.

\subsection{\INDA: Graph Density}

We will exploit the following graph identity~\cite{GraphTheory}
\begin{equation}\label{eq:N_identity_uis}
	N = \frac{\Angles{k}}{\rho}+1, 
\end{equation}
where $\rho=\frac{2|E|}{N(N-1)}$ is the graph density, and $\Angles{k}=\frac{2|E|}{N}$ is the average node degree.

\subsubsection{UIS}
Under UIS, the average node degree $\Angles{k}$ can be easily estimated from our sample~$S$ as 
\begin{equation}\label{eq:av_degree_est_uis}
	\Angles{\est{k}} \ \ = \ \  \frac{\displaystyle \sum_{s\in S}\deg(s)}{n}. 
\end{equation}
The graph density $\rho$ can be interpreted as the probability that two different nodes $u$ and $v$, $u\neq v$,  chosen uniformly at random, are adjacent. 
This can be estimated by inspecting all pairs of different nodes in our sample~$S$, and counting the fraction of them that actually forms edges, \ie
%
\begin{equation}\label{eq:est_rho_uni}
	\est{\rho} \ 
	=\  \frac{\displaystyle\sum_{i<j} 1_{\{\{s_i,s_j\}\in E\}}}  {\displaystyle\sum_{i<j} 1} \ 
	=\ \frac{\displaystyle n^\ind}{n(n-1)/2}.
\end{equation}
%
By plugging \eqn{eq:av_degree_est_uis} and \eqn{eq:est_rho_uni} in \eqn{eq:N_identity_uis}, we then obtain the size estimator
\begin{equation}\label{eq:N_identity_uis_est}
	\est{N}_\UIS^\indA \ 
	=
	\  \frac{\displaystyle (n-1) \cdot \sum_{s\in S}\deg(s) }{2\cdot n^\ind} \  + \ 1.
\end{equation}

\subsubsection{WIS}\label{subsec:theory_WIS}
Under WIS, node $v$ has assigned sampling weight $\w(v)$, which creates a linear bias towards nodes with higher weights. 
We can correct for this bias by applying the Hansen-Hurwitz technique~\cite{HansenHurwitz1943,Gjoka2010}, 
which consists of dividing by $\w(s)$ every term related to $s\in S$. Consequently, the corrected version of \eqn{eq:av_degree_est_uis} becomes~\cite{Kolaczyk2009,Rasti09-RDS,Gjoka2010}: 
\begin{equation}\label{eq:av_degree_est_wis}
	\Angles{\est{k}}\ \ =\ \  \frac{\displaystyle \sum_{s\in S}\frac{\deg(s)}{\w(s)}}{\displaystyle \sum_{s\in S} \frac{1}{\w(s)}}.
\end{equation}
Similarly, we apply a two-point correction~\cite{Kolaczyk2009} to node pairs in \eqn{eq:est_rho_uni}, to obtain the density estimator
\begin{eqnarray}\label{eq:est_rho_wis}
	\est{\rho} &
	=& \ \ \frac
	{\displaystyle\sum_{i<j} \frac{1_{\{\{s_i,s_j\}\in E\}}}{\w(s_i)\w(s_j)}}
	{\displaystyle\sum_{i<j} \frac{1}{\w(s_i)\w(s_j)}}.
\end{eqnarray}
Finally, by plugging \eqn{eq:av_degree_est_wis} and \eqn{eq:est_rho_wis} in \eqn{eq:N_identity_uis}, we obtain the size estimator
\begin{equation}
\label{eq:N1_est_wis} \est{N}_\WIS^\indA \ =\   \ \frac
{\displaystyle\sum_{s\in S}\frac{\deg(s)}{\w(s)} \ \cdot \sum_{i<j} \frac{1}{\w(s_i)\w(s_j)}}  
{\displaystyle\sum_{s \in S} \frac{1}{\w(s)} \cdot \sum_{i<j} \frac{1_{\{\{s_i,s_j\}\in E\}}}{\w(s_i)\w(s_j)}} \ +\ 1. 
\end{equation}

\subsection{\INDB: Arbitrary $A$ and Sample $S$} 
\label{subsec:Approach 2: Two samples}
In this technique, we assume that we have two sets of nodes, $A\subset V$ and $S\subset V$. 
$A$~is an \emph{arbitrary subset} of~$V$ (possibly with repetitions). 
$S$~is an \emph{independence sample} (UIS or WIS), with replacement. 
Our main object of study is the number of cross-collisions $n^\xcol$ between $S$ and $A$, \ie
$$n^\xcol \ =\ \sum_{s\in S}\sum_{a\in A} 1_{\{s=a\}}.$$ 

\subsubsection{UIS} Under UIS, every node in $S$ is selected uniformly at random from all $N$ nodes. Consequently, the probability that $s\in S$ collides with a given $a\in A$ is
$$\Pr(s=a) \ =\  \frac{1}{N}.$$
So the the expected number of collisions is 
$$\Mean{n^\xcol} \ =\ \sum_{s\in S, a\in A}\Pr(s=a) = \frac{|A||S|}{N}.$$
By replacing $\Mean{n^\xcol}$ with the value $n^\xcol$ measured in reality, we obtain the following size estimator:
\begin{equation}\label{eq:N_2samples_uis_est}
	\est{N}_\UIS^\indB \ =\  \frac{|A||S|}{n^\xcol}.
\end{equation}
This resembles capture-recapture~\eqn{eq:capture}, except that here only one phase~($S$) is uniform, and the other one~($A$) is arbitrary. Moreover, we allow for repetitions. 
%

\subsubsection{WIS}
Let us first re-write \eqn{eq:N_2samples_uis_est} as 
\begin{equation}\label{eq:N_2samples_uis_est_full}
\est{N}_\UIS^\indB \ =\ \frac{\displaystyle |A|\cdot \sum_{s\in S}1}{\displaystyle \sum_{s\in S}\sum_{a\in A}1_{\{s=a\}}}.
\end{equation}
Under WIS, 
as in \Sec{subsec:theory_WIS}, the application of the Hansen-Hurwitz estimator to the terms related to $S$ leads to 
%
\begin{equation}\label{eq:WIS_N}
	\est{N}_\WIS^\indB \ =\ \frac{\displaystyle |A|\cdot \sum_{s\in S}\frac{1}{\w(s)}}{\displaystyle \sum_{s\in S}\frac{1}{\w(s)} \sum_{a\in A} 1_{\{s=a\}}}.	
\end{equation}

\subsubsection{How to Choose $A$?}\label{subsec:How to choose A}
Note that in all the derivations above, $A$ is an \emph{arbitrary} set (or multiset) of nodes. The only assumption is that $S$ is drawn independently from~$A$. 
If we happen to have such a set $A$ (\eg from previous measurements or other sources), then we can employ it with our estimators. 
Otherwise (and more conveniently), we can choose $A$ to be all neighbors of nodes in $S$, \ie
\begin{equation}\label{eq:A}
	A=\bigcup_{s'\in S}\mathcal{N}(s') \quad \textrm{(set or multiset)}. 
\end{equation}
%
With this approach, we obtain an $A$ that is:
\begin{itemize}
	\item Relatively large. Indeed, $|A| \approx |S|\cdot\Angles{k}$ under UIS, and $|A| \approx |S|\cdot\Angles{k^2}/\Angles{k}$ under WIS. 
	\item Generally free, \ie with no additional sampling cost. In most graph exploration contexts, we automatically obtain for every $s\in S$ a list of its neighbors $\mathcal{N}(s)$, and can thus employ this list without additional queries.
	\item Almost independent of $S$. Clearly, each node $s\in S$ determines the $\mathcal{N}(s)$ that, in turn, is added to~$A$ (and, consequently, $s$ cannot collide with any node from $\mathcal{N}(s)$). However, $s$ is independent of all the remaining nodes in~$A$, 
	so the dependence of $S$ on $A$ quickly diminishes with growing sample size~$n\eq|S|$. 
	\item Of potentially unknown distribution. For example, when $S$ follows WIS, $A$ depends on graph assortativity~\cite{Newman02}. Similarly, if we discard duplicates, nodes in $A$ may follow a very complex distribution~\cite{Kurant2010}. However, none of these is a problem, because the estimators above accept an arbitrary set~$A$. 	
\end{itemize}

Moreover, under $A$ selected by \eqn{eq:A} (using multiset), we have $n^\xcol \equiv n^\ind$. So, again, we count \emph{edges induced} on the sampled nodes~$S$ (which explains why we use \IND to refer to this category). 


\emph{Set or Multiset?} We can either keep potential node duplicates in $A$ (\ie make $A$ a multiset), 
or discard them (\ie make $A$ a set). Because $A$ can be arbitrary, both of these approaches work well. 
However, we found in simulations that the latter version sometimes performs significantly better
(especially in highly skewed degree distributions), and never worse. 
For this reason, unless explicitly noted, we will henceforth discard all duplicates in~$A$. 

\subsection{\INDA vs. \INDB} 
In all the experiments we conducted both \INDA and \INDB proved asymptotically unbiased. However, \INDB consistently performed better than \INDA in terms of variance. 
This is because \INDA requires two-point correction: a single edge $\{u,v\}\in E$ may have substantial weight in \eqn{eq:N1_est_wis} if $\w(u)$ and $\w(v)$ are small (\ie exactly when $u$ and $v$ are rarely sampled), increasing the variance of the estimator. 
In contrast, \INDB uses only one-point corrections, which makes it more robust. 

For this reason, and to improve the paper's readability, we will henceforth use only the \INDB technique, and we will refer to it simply as \IND.


%
%

\section{Dependence Reduction for RW}\label{sec:random_walk}

Both UIS and WIS select nodes independently. In practice, this can be difficult or impossible to achieve (see our discussion in Footnote~\ref{footnote:UIS}). 
In contrast, one can often perform a random walk (RW), as commonly done in WWW~\cite{Henzinger2000}, P2P networks~\cite{Stutzbach2006-unbiased-p2p,Rasti09-RDS, Gkantsidis2004} and  OSNs~\cite{Twitter08,Rasti2008,Gjoka2010,Mohaisen2010,Lu2012}. 
In an undirected, connected and acyclic graph, RW visits node $v$ at a given step with probability proportional to its degree~$\deg(v)$. Therefore, one could be tempted to set $\w(v)=\deg(v)$ and apply directly the WIS estimators from~\Sec{sec:Graph size estimation based on induced subgraph}. 

\begin{figure}
\psfrag{IND}[l][c][1]{a) \NODE (state of the art)}
\centering
\includegraphics[width=0.48\textwidth]{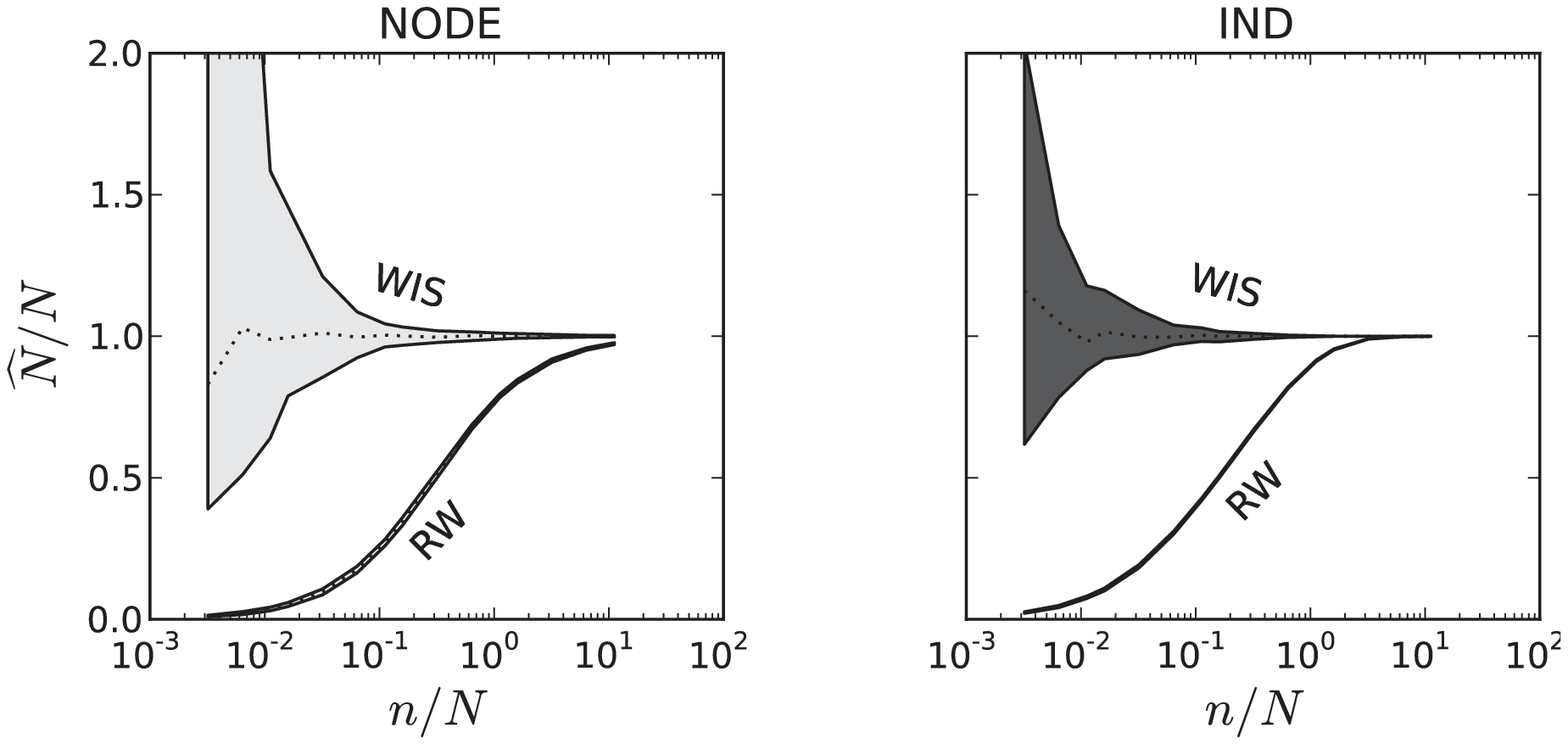}
\caption{The estimated size $\est{N}/N$, as a function of relative sample length~$n/N$,  
for \texttt{p2p-Gnutella31} graph (see \Tab{table:topologies}); other graphs yield analogous results. 
}
\label{fig:RW_vs_n}
\end{figure}

Unfortunately, as we demonstrate in \Fig{fig:RW_vs_n}, this approach fails. 
Indeed, under RW, the estimate $\est{N}$ can be arbitrarily small for small~$n$. 
%
%
This effect fades away for much larger~$n$, say for $n>N$. However, taking so large sample is of course impractical - the central goal of sampling is to estimate some properties based on a relatively small sample, \ie where $n\ll N$.

The WIS estimators fed directly by RW samples perform poorly because of the strong \emph{dependence} between consecutive draws. 
Assume, for example, that our RW sample consists of just three nodes, \ie $S=[s_1,s_2,s_3]$. 
Under RW, $s_1$ and $s_3$ collide ($s_1=s_3$) with probability equal to $1/\deg(s_2)$.  In contrast, under WIS, this probability may be arbitrarily close to 0 (for $N\rightarrow\infty$). 
So RW experiences increased number of collisions~$n^\col$, which leads to the underestimation of~$N$.

Clearly, in order to apply the WIS graph size estimators to a RW sample, we have to reduce the dependence created by the underlying Markov chain. Below, we describe one simple dependence reduction technique used in the MCMC literature, and then we propose significantly more efficient techniques.

\subsection{\SimpleThinning} \label{subsec:Simple thinning}
The authors of \cite{Katzir2011}  reduce RW dependence by taking every $\theta$th sample from~$S$, where $\theta$ is a \emph{thinning} parameter. 
The resulting subsample 
\begin{equation}\label{eq:RW-single}
	S'\ =\ [s_{1},s_{1+\theta},s_{1+2\theta},\ldots]
\end{equation}
is then fed to \eqn{eq:est(N) WIS} to obtain a size estimate. 

This approach has several drawbacks. 
First of all, $(\theta-1)/\theta$ samples are dropped, which is a clear waste. 
Second, as we will see in \Sec{sec:results}, it may be very challenging (often impossible) to find the optimal value of~$\theta$.  

\subsection{\ShiftedThinning}\label{subsec:Thinning with shifting}

\SimpleThinning can be easily improved by observing that, rather than one, we obtain $\theta$ different subsamples:
\begin{equation}\label{eq:RW-multiple}
  S'_k\ =\ [s_{1+k},\ s_{1+k+\theta},\ s_{1+k+2\theta},\ldots], \ \ \ k=0\ldots \theta-1. 
\end{equation}
One way to exploit all these subsamples is to apply a size estimator to each of them, creating $\theta$ different estimates~$\est{N}(S'_k)$. We may take the mean or median of them as our final result, \eg
\begin{equation}
\label{eq:N_aggregated_naive}
\est{N}_{aggregated}\ \ = \ \ \frac{1}{\theta} \cdot \sum_{i=0}^{\theta-1} \est{N}(S'_k). 
\end{equation}
However, this can be problematic \eg if for some or many $k$, we have $\est{N}(S'_k)\eq\infty$. 
Instead, we propose to aggregate the $\theta$ estimates is by applying
\begin{equation}
\label{eq:N_aggregated}
\est{N}_{aggregated}\ \ = \ \ \frac{\displaystyle \sum_{i=0}^{\theta-1} \textrm{numerator}(\est{N}(S'_k))}{\displaystyle \sum_{i=0}^{\theta-1} \textrm{denominator}(\est{N}(S'_k))},	
\end{equation}
where $\textrm{numerator}(\est{N}(S'_k))$ is the numerator of the estimator $\est{N}(S'_k)$; analogously for the denominator. 
This approach avoids the $\est{N}(S'_k)\eq\infty$ problem and performs (in simulations) consistently better than \eqn{eq:N_aggregated_naive}. We will henceforth use \ShiftedThinning to refer to \eqn{eq:N_aggregated}.

%
%
%
%
%
%

\subsection{\Margin}\label{subsec:Margin-based solution}

We propose yet another approach to reduce dependence in a RW sample. Our main idea is to ignore the information brought by pairs of nodes that are less than $m$ samples away. This should leave us with pairs of independently selected nodes only. To achieve this, we must in some cases modify our estimator. 

Applying this idea to NODE (\eqn{eq:est(N) WIS}) is rather straightforward, and leads to 
\begin{eqnarray}
\label{eq:est_N WWW_margin}	
\est{N}_\RW^{\scriptscriptstyle\textrm{NODE}}  &=& \ \frac
{\displaystyle \sum_{i,j} \frac{\w(s_i)}{\w(s_j)} \cdot 1_{\{|j-i|>m} }
{\displaystyle \sum_{i,j} 1_{\{s_i=s_j\}} \cdot 1_{\{|j-i|>m} }.
\end{eqnarray}

In contrast, the \IND estimator (\eqn{eq:WIS_N} with multiset~$A$) require some additional transformations, as follows. 
First, note that 
$$\sum_{a\in A} f(a) \equiv \sum_{s\in S} \sum_{a\in \mathcal{N}(s)} f(a), \textrm{\quad and \quad} |A| = \sum_{s\in S} \deg(s)$$
Consequently, \eqn{eq:WIS_N} can be rewritten as 
\begin{eqnarray}
\nonumber	\est{N}_\WIS^\ind 
&=& 
 \frac
{\displaystyle \left(\sum_{i} \deg(s_i)\right)  \left(\sum_{j} \frac{1}{\w(s_j)}\right)}
{\displaystyle \sum_{i} \frac{1}{\w(s_i)} \sum_{j}\sum_{a\in \mathcal{N}(s_j)}{1_{\{s_i=a\}}}} \\ 
\nonumber &=&
 \frac
{\displaystyle \left(\sum_{i} \deg(s_i)\right)  \left(\sum_{j} \frac{1}{\w(s_j)}\right)}
{\displaystyle \sum_{i} \frac{1}{\w(s_i)} \sum_{j}{1_{\{s_i\in \mathcal{N}(s_j)\}}}} \\ 
\nonumber &=&
\frac
{\displaystyle \sum_{i,j} \frac{\deg(s_i)}{\w(s_j)}}
{\displaystyle \sum_{i,j} \frac{1_{\{s_i\in \mathcal{N}(s_j)\}}}{\w(s_i)}}. 
\end{eqnarray}
Now, it is easy to exclude the pairs of nodes lying within $m$ hops, \ie
\begin{equation}
\label{eq:N_RW__IND}
\est{N}_\RW^\ind = 
\frac
{\displaystyle \sum_{i,j} \frac{\deg(s_i)}{\w(s_j)} \cdot 1_{\{|j-i|>m\}}}
{\displaystyle \sum_{i,j} \frac{1_{\{s_i\in \mathcal{N}(s_j)\}}}{\w(s_i)}\cdot 1_{\{|j-i|>m\}}}.	
\end{equation}
\eqn{eq:N_RW__IND} interprets $A$ as a multiset, which is different from the set version that we suggested in \Sec{subsec:How to choose A}. Although we omit it here (for brevity), we implemented the latter and use in in the evaluation.


\smallskip

Finally, we would like to note that \Margin naturally fits the sampling strategies using multiple independent RWs (as \eg in \cite{Gjoka2010}). Indeed, it is enough to replace in \eqn{eq:est_N WWW_margin}	and \eqn{eq:N_RW__IND} every term $ 1_{\{|j-i|>m\}}$ with $ 1_{\{\textrm{walker}(i)\neq \textrm{walker}(j)\}}$, where $\textrm{walker}(i)$ is the walker that contains sample~$s_i$. In other words, we consider only the node pairs where nodes come from different walks, and are thus independent. Note that the resulting estimators have no explicit parameter~$m$. 

\subsection{Comparison}\label{subsec:Comparison of decorrelation techniques}

To compare our dependence reduction techniques, first note that the main information exploited by our estimators lies in \emph{pairs of sampled nodes}. 
For example, \eqn{eq:est_N WWW} uses \eqn{eq: n^col} that explicitly considers all node pairs and counts their collisions. 
Similarly, in denominator of \eqn{eq:WIS_N} with $A$ constructed by \eqn{eq:A}, we count collisions between a sampled node~$s_i$ and the neighbors of another sampled node~$s_j$. 
Consequently, the efficiency of an estimator grows with the number of node pairs it considers. 

In the entire sample~$S$, $|S|=n$, we have $n(n-1) \cong n^2$ node pairs.\footnote{In this simple calculation, we count separately pairs $(s_i,s_j)$ and $(s_j,s_i)$. Indeed, these two pairs bring different information to  \eqn{eq:WIS_N} with \eqn{eq:A}.}
UIS and WIS estimators make use of all of them. 
In contrast, the RW dependence reduction techniques proposed in \Sec{subsec:Simple thinning}-\Sec{subsec:Margin-based solution} may significantly reduce the number of considered pairs, as follows (see \Tab{table:number of node pairs used}). 

\SimpleThinning uses only $\frac{n}{\theta}$ nodes, which results in $\frac{n^2}{\theta^2}$ node pairs. 
Analogously, \ShiftedThinning uses $\frac{n^2}{\theta^2}$ node pairs for each $k=0\ldots\theta-1$, which results in the total of $\frac{n^2}{\theta}$ node pairs. 
Finally, from all $ n^2$ node pairs, \Margin drops $2m$ pairs in the neighborhood of each node. Since there are $n$ such nodes, we keep $n^2\m 2nm=n(n\m 2m)$ node pairs. 

Both $\theta$ and $m$ signify the same notion---the number of Markov chain steps such that the dependence between RW samples becomes negligible. In typical networks this happens for $\theta(=m)$ in the order of tens to hundreds~\cite{Mohaisen2010}. 
Consequently, $\frac{n^2}{\theta^2} \ll \frac{n^2}{\theta} \ll  n(n-2m)$, and we may expect the \Margin to perform best.

\begin{table}
\centering{
\begin{tabular}{  c | c}

Dependence Reduction Method & Node Pairs\\
\hline 
\SimpleThinning & $\frac{n^2}{\theta^2}$\\
\ShiftedThinning & $\frac{n^2}{\theta}$\\
\Margin & $n(n-2m)$\\
\end{tabular}    
}
	\caption{Approximate number of node pairs exploited by each of the dependence reduction techniques.}
	\label{table:number of node pairs used}
\end{table}

\section{Implementation issues} \label{sec:Implementation issues}
A straightforward, naive implementation of the above estimators can easily lead to $O(n^2)$ time complexity, where $n\eq|S|$ is the sample size. This is the case, for example, for the sum 
$$\sum_{i<j} \frac{1}{\w(s_i)\w(s_j)}$$ 
in \eqn{eq:N1_est_wis}. Although not a problem for small samples, $O(n^2)$ may become an issue, say, for $n>100K$. Because our real-life samples are often significantly larger (\eg we sampled millions of Facebook nodes), we had to look for more efficient implementations of our estimators. Fortunately, all of them can be rewritten to use only $O(n)$ time complexity. For example, one can easily show that the above sum is equal to
$$\frac{1}{2}\cdot \left(\left(\sum_{i}\frac{1}{\w(s_i)}\right)^2 
- 
\sum_{i}\frac{1}{(\w(s_i))^2}\right).$$

Things become more complicated for RW-targeted estimators in \Sec{sec:random_walk}. Here, the corresponding sums are much more interdependent (especially when we use the ``set'' version of~\eqn{eq:A}), and thus difficult to separate. However, even in this case, the time complexity can be kept linear, with the help of some auxiliary dedicated data structures.

Our \texttt{python} implementation available at \cite{WWW_N_est} guarantees $O(n)$ for all estimators derived in this paper.  

\section{Performance Evaluation}\label{sec:results}

In this section, we evaluate the \NODE and \IND estimators under three sampling techniques UIS, WIS and RW. We apply them to a wide spectrum of real-life fully known topologies (\Sec{subsec:offline}) and well as to several samples of Facebook (\Sec{subsec:online}). 
\Tab{table:Estimators used in simulations} summarizes the concrete estimators we used in this study.

\begin{table}
\centering{
\small{
\begin{tabular}{@{} r @{}| c | c|}
 & NODE & IND \\
 \hline 
 UIS & \cellcolor[gray]{0.9} \eqn{eq:est(N) UNI collisions} & \cellcolor[gray]{0.45} \eqn{eq:N_2samples_uis_est} \\
 WIS & \cellcolor[gray]{0.9} \eqn{eq:est(N) WIS} & \cellcolor[gray]{0.45} \eqn{eq:WIS_N}\\
 RW, \SimpleThinning & \cellcolor[gray]{0.9}  \eqn{eq:est(N) WIS}\!+\!\eqn{eq:RW-single}
  & \cellcolor[gray]{0.45} \eqn{eq:WIS_N}\!+\!\eqn{eq:RW-single}  \\
 RW, \ShiftedThinning  & \cellcolor[gray]{0.7}\eqn{eq:est(N) WIS}
 \!+\!\eqn{eq:N_aggregated}  & \cellcolor[gray]{0.45} \eqn{eq:WIS_N}
 \!+\!\eqn{eq:N_aggregated}\\
 RW, \Margin & \cellcolor[gray]{0.9}  \eqn{eq:est_N WWW_margin}	 & \cellcolor[gray]{0.45} \eqn{eq:N_RW__IND} \\
\hline 	  
\end{tabular}    
}
}
\caption{Estimators used in simulations. 
The shades of gray correspond to those used in \Fig{fig:simulations_WIS}, \Fig{fig:simulations_RW} and \Fig{fig:FB}.
	}
	\label{table:Estimators used in simulations}
\end{table}

\subsection{Fully Known Topologies}\label{subsec:offline}

We first evaluate our estimators on fully-known topologies, which allows us to compare the results directly with the ground-truth graph size. We used 19 real-life topologies coming from various fields, with up to millions of nodes and tens of millions of edges. They are summarized in~\Tab{table:topologies}.

\subsubsection{UIS}
In \Fig{fig:simulations_WIS}(a), we present the simulation results under UIS sampling. 
First of all, we observe that both \NODE and \IND converge to the correct value (1.0 on y-axis) as sample size $n$ grows. 
%
Second, in all cases, \IND outperforms \NODE. 
For example, for \texttt{Berkeley13}, the \IND estimator with sample size $n\eq200$ performs similarly to \NODE with $n\eq2000$. This means that \IND reduces the sampling cost by 90\%, compared to \NODE. 
This advantage of \IND over \NODE depends on many factors, in particular on mean degree. Indeed, all graphs with high average node degree $\bar{k}$ experience several-fold improvement under \IND. 
In contrast, for graphs with $\bar{k}<5$ (\eg ``email-EUAll'' or ``roadNet-PA''), the difference is much less pronounced.


\subsubsection{WIS}
\begin{figure*}
\subfigure[UIS]{
\includegraphics[width=1.0\textwidth]{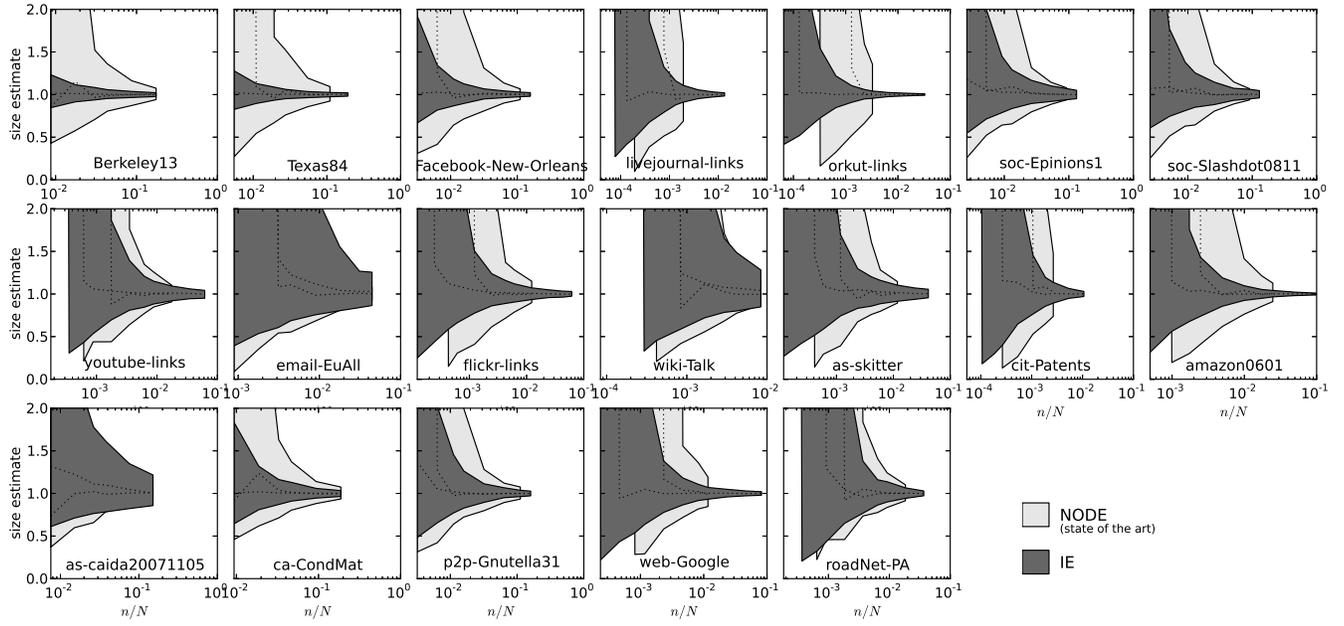}
}
\subfigure[WIS]{
\includegraphics[width=1.0\textwidth]{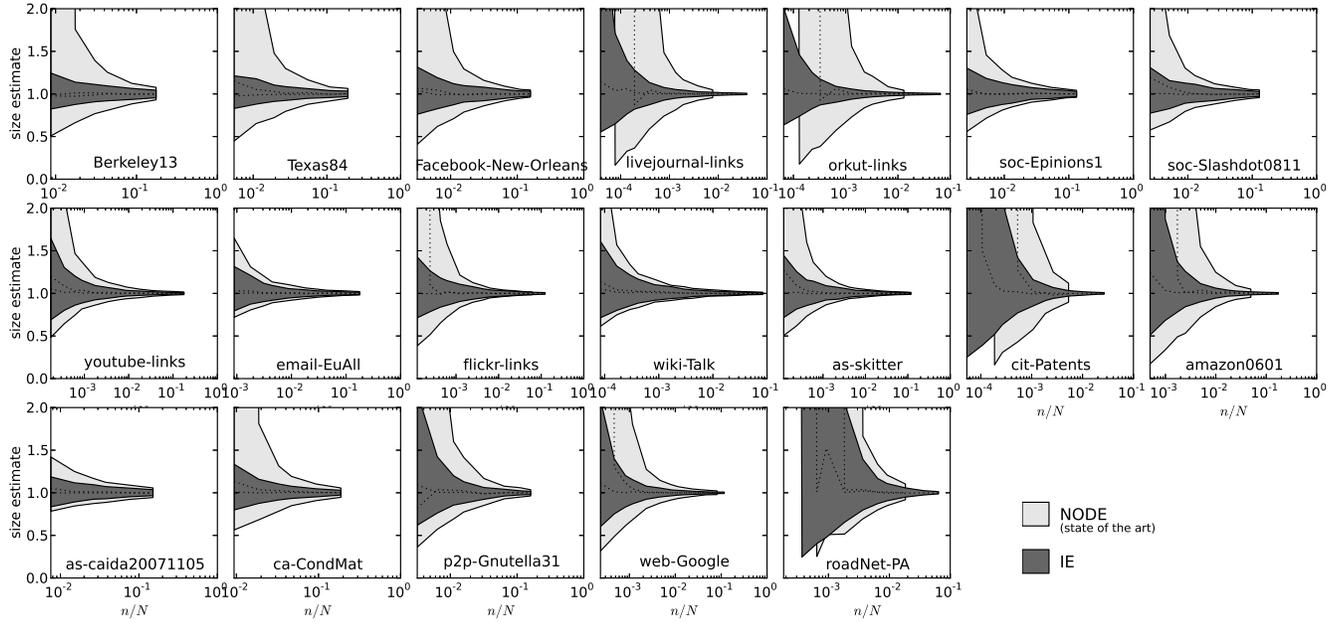}
}
\caption{[UIS and WIS] The estimated size $\est{N}$ relative to the real size~$N$ for 19 real-life fully known topologies, as a function of relative sample length $n/N$. We use two sampling techniques: UIS~(a) and WIS~(b) (under WIS, nodes were selected from the stationary distribution of RW). 
We consider two estimation techniques: \NODE~(light gray) and \IND~(dark gray). For every sample length~$n$, we performed 500 experiments. 
The grey regions cover the 500 results from 10th percentile to 90th percentile, with the median set in dotted line.
}
\label{fig:simulations_WIS}
\end{figure*}

Under WIS (see \Fig{fig:simulations_WIS}(b)), the efficiency of both methods improves, especially for sparser topologies. 
This is because heterogeneous sampling weights result in more collisions in the sample, which, in turn, gives the estimators more information to exploit. 
(The same phenomenon has already been observed for \NODE in~\cite{Katzir2011}). 
However, the relative advantage of \IND over \NODE remains roughly the same as under UIS, and is, again, primarily determined by mean degree.  

\begin{table}
\small{
\begin{tabular}{@{} l | r @{}| r @{}| r @{}| r @{} }
name &     nodes \ &    edges \ &  $\Angles{\deg}\ $ & $\frac{\Angles{\deg^2}}{\Angles{\deg}}$ \\
\hline 
           Berkeley13~\cite{Traud2011}&       22K &      852K &  74.4 & 167.0 \\
             Texas84~\cite{Traud2011} &       36K &     1\,590K &  87.5 & 212.1 \\
Facebook-New-Orleans~\cite{Viswanath2009} &       63K &      816K &  25.8 &  88.1\\ 
   livejournal-links~\cite{Mislove2007} &     5\,189K &    48\,688K &  18.8 & 155.4 \\
orkut-links~\cite{Mislove2007} &     3\,072K&   117\,185K &  76.3 & 390.3 \\
       soc-Epinions1~\cite{Richardson2003}&       75K &      405K &  10.7 & 183.9 \\
    soc-Slashdot0811~\cite{Leskovec2009} &       77K &      469K &  12.1 & 147.0 \\    
			 youtube-links~\cite{Mislove2007} &     1\,134K &     2\,987K &   5.3 & 494.5 \\
         email-EuAll~\cite{Leskovec2007} &      224K &      339K &   3.0 & 567.6 \\
         flickr-links~\cite{Mislove2007} &     1\,624K &    15\,476K &  19.0 & 949.2\\
        wiki-Talk~\cite{Leskovec2010} &     2\,388K &     4\,656K &   3.9 & 2\,705.4\\  
          as-skitter~\cite{Leskovec05_Forest_Fire} &     1\,694K &    11\,094K &  13.1 & 1\,445.1 \\
				 cit-Patents~\cite{Leskovec05_Forest_Fire} &     3\,764K &    16\,511K &   8.8 &  21.3 \\
          amazon0601~\cite{Leskovec2007a} &      403K &     2\,443K &  12.1 &  30.6 \\
    as-caida20071105~\cite{Leskovec05_Forest_Fire} &       26K &       53K &   4.0 & 280.2 \\
          ca-CondMat~\cite{Leskovec2007} &       21K &       91K &   8.5 &  22.5 \\         
      p2p-Gnutella31~\cite{Leskovec2007} &       62K &      147K &   4.7 &  11.6 \\
          web-Google~\cite{Leskovec2009}&      855K &     4\,291K &  10.0 & 170.4 \\
          roadNet-PA~\cite{Leskovec2009}&     1\,087K &     1\,541K &   2.8 &   3.2 
\end{tabular}    
}
	\caption{Topologies used in offline simulations in \Sec{subsec:offline}.  $\Angles{\deg}$ is the average node degree, $\Angles{\deg^2}$ is average squared node degree. High value of $\Angles{\deg^2}/\Angles{\deg}$ compared to $\Angles{\deg}$ indicates a highly heterogeneous node degree distribution.}
	\label{table:topologies}
\end{table}

\begin{figure*}
\subfigure[\SimpleThinning~(\Sec{subsec:Simple thinning}) and \ShiftedThinning~(\Sec{subsec:Thinning with shifting}). On x-axis, we vary the thinning parameter~$\theta$. ]
{
\includegraphics[width=1.0\textwidth]{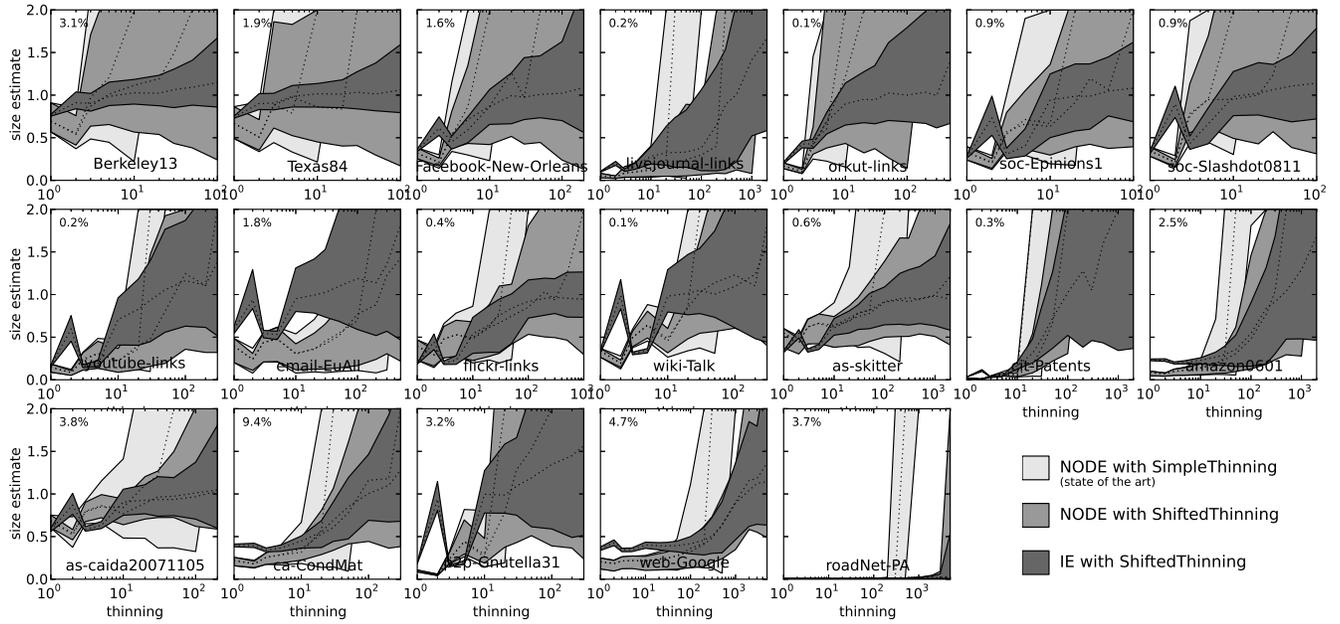}
}
\subfigure[\Margin~(\Sec{subsec:Margin-based solution})]{
\includegraphics[width=1.0\textwidth]{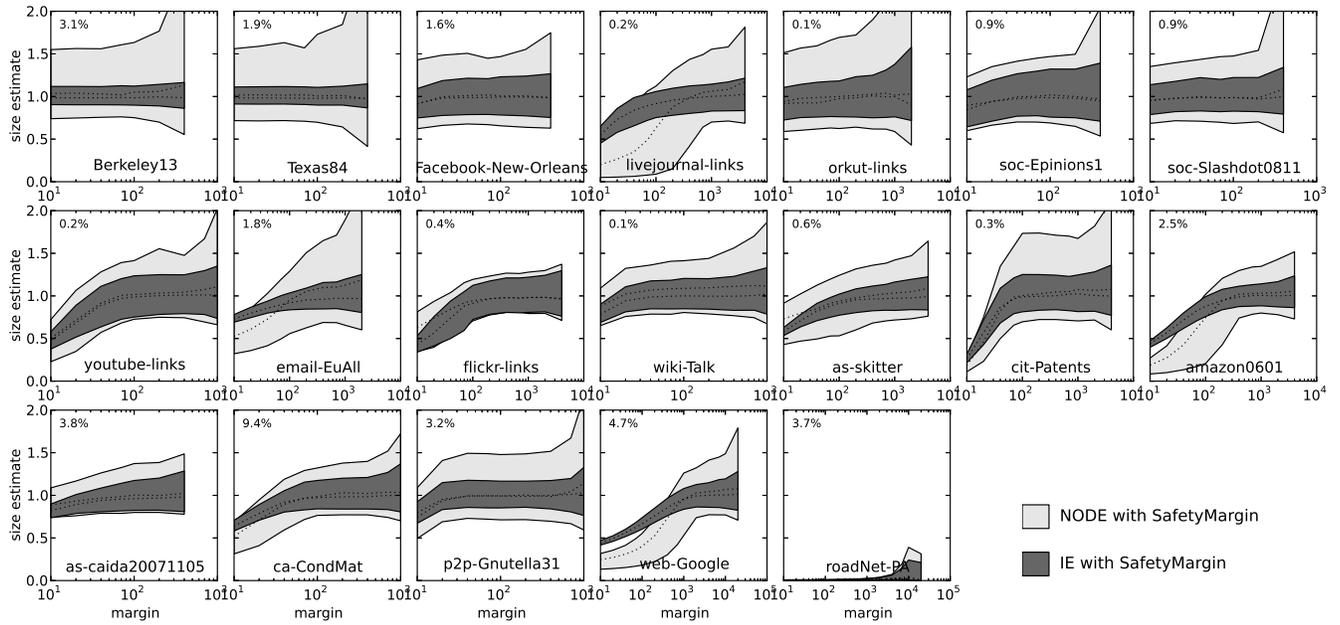}
}
\caption{[RW] The estimated size $\est{N}$ relative to the real size~$N$, for 19 real-life fully known topologies sampled with RW of relative length  $n/N$ given in top left corners. 
We use two RW dependence reduction techniques: Thinning~(a) and \Margin~(b). 
Under each, we test two estimation techniques: \NODE (light and medium grey) and \IND (dark grey). 
%
\quad For every topology, we performed 500 experiments. 
The grey regions cover the 500 results from 10th percentile to 90th percentile, with the median set in dotted line. 
}
\label{fig:simulations_RW}
\end{figure*}

\begin{figure*}
\includegraphics[width=1.0\textwidth]{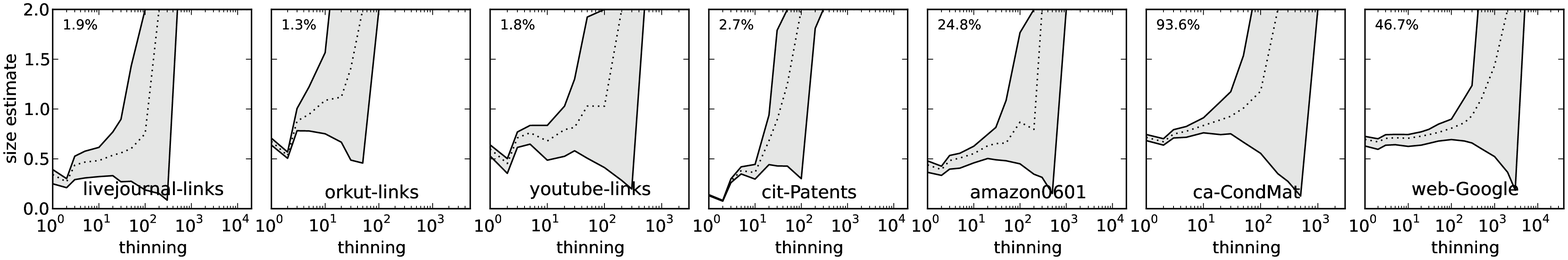}
\caption{[RW] 
NODE with \SimpleThinning (state of the art) for RW with ten times larger sampling budget than in~\Fig{fig:simulations_RW}(a,b), in 7 example topologies.}
\label{fig:sim_thinning_10times_longer}
\end{figure*}

\subsubsection{RW}

In \Fig{fig:simulations_RW}, we present the simulation results for RW sampling, with two dependence reduction techniques: Thinning~(a) and \Margin~(b). For each topology, we fix the sampling budget, and we vary the thinning parameter~$\theta$ in (a) and the margin~$m$ in (b). 

\paragraph{Thinning}
We analyze Thinning in \Fig{fig:simulations_RW}(a). 
In general, \IND with \ShiftedThinning outperforms \NODE with \ShiftedThinning, which in turn outperforms \NODE with \SimpleThinning. This is in agreement with our analysis in \Sec{subsec:Comparison of decorrelation techniques}. 
All versions of thinning follow the same general pattern with two or three regimes of $\theta$:
\begin{enumerate}
	\item \emph{Underestimation}: For $\theta\rightarrow 1$, the thinning is too weak, and the RW dependence results in a systematic underestimating of size. 

	\item \emph{Flattening} (not guaranteed): In some topologies, for some range of~$\theta$, the estimate stabilizes around the true value, with acceptable variance. 


	\item \emph{Overestimation}: For $\theta\rightarrow\infty$, we observe no collisions within the thinned samples and thus our estimate is often $\est{N}\eq\infty$. This effect can be easily observed for \NODE where many plots shoot upwards for larger~$\theta$.
\end{enumerate}
Only if \emph{Flattening} is present (and well pronounced) can one try to interpret the results and estimate the graph size. 
In \Fig{fig:simulations_RW}(a), this is the case, say, for \texttt{Berkeley13}, \texttt{Texas84}, \texttt{orkut-links} and \texttt{wiki-Talk} under \IND (although this assessment is very subjective in nature). 
In all other cases, including all NODE cases, \emph{Flattening} does not occur, making them essentially impossible to interpret.

\paragraph{\Margin} In contrast, the \Margin performs very well, as shown in~\Fig{fig:simulations_RW}(b). 
Here, we can observe the same three regimes (\emph{Underestimation}, \emph{Flattening}, \emph{Overestimation}), as under Thinning. 
However, now \emph{Flattening} is very well pronounced: it spans a wide range of margins~$m$, yields a relatively small variance, and concentrates around the true value. 
This makes the results much easier to interpret. 

The only exception is \texttt{roadNet-PA} (last topology), where all of our RW estimators fail miserably. 
This is probably because \texttt{roadNet-PA} represents a road network, which is typically a lattice-like, almost planar graph, with a very large diameter (here diam=782). 
Consequently, the mixing time of RW (and thus the desired margin~$m$) is very large, possibly larger than the sample sizes we tested. 
Indeed, under the absence of RW dependence, our estimators perform well, as presented in \Fig{fig:simulations_WIS}(a,b).

\paragraph{Comparison with State-of-the-art Techniques}
To date, the state of the art has been NODE with \SimpleThinning~\cite{Katzir2011}. 
We show its performance in \Fig{fig:sim_thinning_10times_longer}, for RW ten times longer than in~\Fig{fig:simulations_RW}(a,b).  
None of the presented plots enters the \emph{Flattening} regime, which makes the estimation impossible (the same holds for NODE with \ShiftedThinning, not shown). 
 In contrast, \IND with \Margin in \Fig{fig:simulations_RW}(b), performed very well even for RW samples of 1/10th the length. This means that, compared to the state of the art, 
 our techniques achieved here \emph{more than 10-fold reduction in sampling cost}.

Interestingly, a closer comparison of \Fig{fig:simulations_WIS}(a) with \Fig{fig:simulations_RW}(b) reveals that 
\IND with \Margin applied to RW (thus a highly interdependent and challenging sample) 
is often better than NODE applied to UIS (independence sample). 

\subsection{Online experiments}\label{subsec:online} 

\begin{figure*}
\centering
\includegraphics[width=0.999\textwidth]{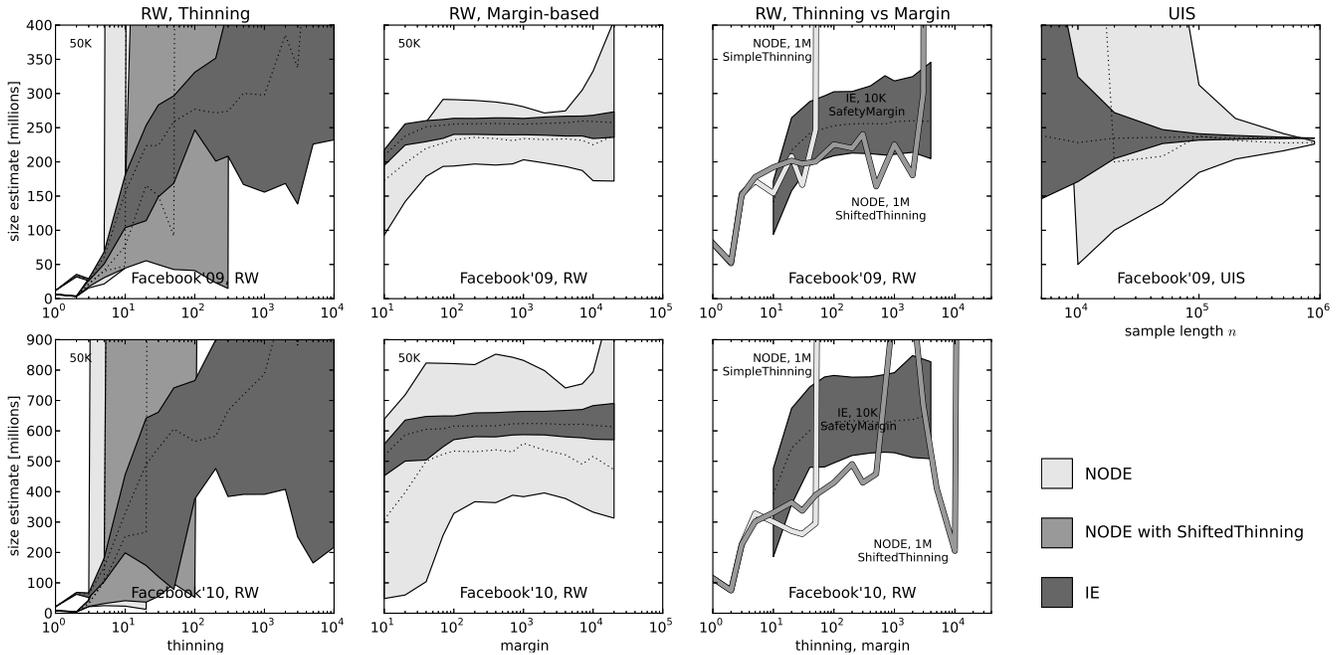}
\caption{Online experiments on Facebook'09 (top) and Facebook'10 (bottom). 
We use RW and UIS (for '09 only) sampling techniques, and consider two size estimators: \NODE~(light and medium gray) and \IND~(dark gray). 
\quad 
The grey regions cover the results from 10th percentile to 90th percentile, with the median set in dotted line. 
%
}
\label{fig:FB}
\end{figure*}

Finally, we test our techniques in online experiments on Facebook, where the entire topology is unknown to us. 
We use samples collected in two different periods of time:\\
${}$\ \textbf{Facebook'09}~\cite{Gjoka2010}: RW and UIS, with 1M users each.\\
${}$\ \textbf{Facebook'10}~\cite{Kurant2011_SWRW}: RW covering 1M users.\\
We show the results in~\Fig{fig:FB}.

\subsubsection{UIS}

In the top-right plot in~\Fig{fig:FB}, we estimate the size of Facebook'09 based on a UIS sample of its users~\cite{Gjoka2010}, as a function of the sampling length~$n$. 
Both \NODE and \IND return values concentrated around $N=240M$ (already reported in~\cite{Katzir2011} with \NODE), which is in agreement with what Facebook claimed at that time. 
However, the \NODE values are much more dispersed than those of \IND. 
For example, for $n=100K$, \NODE 10-90 percentiles correspond to \IND at less than $n=10K$. This means that with \IND we need ten times fewer samples to achieve the \NODE's accuracy, which translates into \emph{10-fold reduction in sampling cost}. 

We should note, however, that a UIS sample of nodes is rarely available. For example, the UIS sample used above was obtained through rejection-sampling of the entire 32-bit userID space. Soon afterwards, Facebook moved to a 64-bit space, which makes this approach completely impractical. In this case, one has to use other methods, such as RW.  Unlike existing methods, our techniques continue to be useful in this case.

\subsubsection{RW}
All the remaining plots in~\Fig{fig:FB} are generated based on RW samples of Facebook nodes, with different dependence reduction techniques. 

The left-most column uses Thinning. Similarly to (most of) \Fig{fig:simulations_RW}(a), the estimates do not stabilize with the thinning parameter $\theta$, which makes the results practically impossible to interpret. 

In contrast, \Margin applied to the same RW samples (second column in~\Fig{fig:FB}) performs very well and leads to good and concentrated size estimates. 

The third column of~\Fig{fig:FB}, is our attempt to compare the efficiency of our estimators. 
To this end, we applied NODE with Thinning to the entire RW sample (with 1M=1 million nodes), which resulted in a single estimate per $\theta$, represented by the light- and medium-grey lines. 
Next, we applied \IND with \Margin to one hundred 10K-long chunks of our RW sample (dark-grey region). 
In both datasets, the state-of-the-art solution, \ie NODE with \SimpleThinning, performed badly.\footnote{Although \ShiftedThinning improves the results, they are still impossible to interpret, especially under Facebook'10.}
In contrast, \IND with \Margin leads to very reasonable estimates. Because the latter uses 100 times fewer node samples, we conclude that in RW sampling of Facebook, 
our techniques lead to at least \emph{100-fold reduction in sampling cost}.

\section{Conclusion and Future Work}\label{sec:conclusion}

In this paper, we began by introducing \IND, an efficient technique to estimate the size of a graph, based on an independence sample (uniform or not) of its nodes. 
In many practical applications, however, independence sampling is not possible, but it is relatively easy to perform a Random Walk (RW) in the graph. Because of the strong dependence between consecutive nodes in an RW sample, neither standard estimators nor \IND can use such data without adjustment. 
To address this problem, we introduced \Margin \ - a technique that corrects the estimators for dependence in RW samples, and is applicable to both \IND and already extant estimation methods.

We evaluated our techniques in simulations on a wide range of fully known real-life topologies, and on several samples of Facebook (confirming the officially announced number of users). We found that, for the same estimation error, \IND with \Margin often requires 10+ times fewer samples than the state-of-the-art solutions. In particular, for Facebook, we observed more than 100-fold reduction in sampling cost. 

A \texttt{python} implementation of all estimators used in this paper, optimized to guarantee $O(n)$ time complexity, is available at \cite{WWW_N_est}.

In future work, we plan to study ideas that can further improve the efficiency of our estimators. 
For example, one can try to use all neighbors of the sampled nodes, rather than the sampled neighbors only, or to 
combine \NODE and \IND together.  
Another challenge is to extend these results to directed graphs, for which RW sampling weights are harder to obtain than in the undirected case; the theory here is unchanged, but the challenges associated with RW sampling per se are significant.
Finally, we plan to study how the \Margin applies to other problems, \eg to the estimation of graph clustering and assortativity.  

\bibliographystyle{abbrv}
\bibliography{OSN_Sampling,predictive_seeding}


\end{document}